\documentclass[12pt,preprint]{aastex}

\lefthead{Bekki, Couch, Drinkwater, \& Shioya}
\righthead{Nuclear cannibalism}

\begin{document}
\title{Cluster cannibalism and scaling relations of galactic stellar nuclei}

\author{Kenji Bekki,  Warrick J. Couch} 
\affil{
School of Physics, University of New South Wales, Sydney 2052, Australia}

\author{Michael  J. Drinkwater}
\affil{Department of Physics, University of Queensland, Queensland 4072, 
Australia}

\and

\author{Yasuhiro Shioya}
\affil{Astronomical Institute, Tohoku University, Sendai, 980-8578, Japan}

\begin{abstract}

Recently, very massive compact stellar systems have been discovered in the
intracluster regions of galaxy clusters and in the nuclear regions
of late-type disk galaxies. 
It is unclear how these compact stellar systems -- known as ``ultra-compact 
dwarf'' (UCD) galaxies or ``nuclear clusters'' (NCs) -- form and evolve.  
By adopting a formation scenario where these stellar systems are the product of 
multiple merging of star clusters in the central regions of galaxies,
we investigate, numerically, their physical properties. 
We find that physical correlations between velocity dispersion, luminosity, 
effective radius, and average surface brightness in the stellar merger remnants 
are quite different from those observed in globular clusters. 
We also find that the remnants have triaxial shapes with or without figure 
rotation, and these shapes and their kinematics depend strongly on the initial 
number and distribution of the progenitor clusters.
These  specific predictions can be compared with the corresponding
results of ongoing and future observations of UCDs and NCs, 
thereby providing a better understanding of the origin of these enigmatic 
objects. 

\end{abstract}

\keywords{galaxies: nuclei --- galaxies: dwarfs --- globular clusters:general --
galaxies: star clusters} 

\section{Introduction}

A new type of sub-luminous and extremely compact ``dwarf galaxy'' has
been recently discovered in an `all-object'  spectroscopic survey
centered on the Fornax Cluster (Drinkwater et al. 2000).
These ``dwarf galaxies'', which are members of the Fornax Cluster,
have intrinsic sizes of less than 100 pc and absolute
$B$ band magnitude  ranging from $-13$ to $-11$ mag
and are thus called ``ultra-compact dwarf'' (UCD) galaxies.
Although these UCDs are suggested to originate from stellar nuclei
of bright nucleated dwarf galaxies 
(Drinkwater 2000; Bekki et al. 2001, 2003),
it is unclear how such massive nuclei are formed in the central region 
of dwarf galaxies.

Recent {\it Hubble Space Telescope} (HST) photometric observations
have  discovered very luminous
nuclear clusters (NCs), with $I$-band absolute magnitudes ($M_{\rm I}$)
ranging from $-8$ to $-14$ mag,
in the central regions of late-type spirals 
(Phillips et al. 1996; Carollo et al. 1998; Matthews et al. 1999;
B$\ddot{\rm o}$ker et al. 2002, 2004a,b).
The observation that some of the bright NCs are quite massive -- that is their 
luminosity does not derive from a small number of hot young stars --   
has raised the question as to how such massive NCs can be formed in the central 
regions of late-type spiral galaxies 
(B$\ddot{\rm o}$ker et al. 2004a,b). 

One formation scenario for these very massive star clusters 
(hereafter referred to as ``VMSCs'') is that ordinary 
star clusters, which can quickly spiral into the nuclear regions of galaxies
due to dynamical friction, 
merge with one another to form a single VMSC (i.e., galactic stellar nuclei; 
Tremaine et al. 1975).
The physical properties of VMSCs formed in this way, 
however, have not been theoretically/numerically investigated extensively 
(e.g., Fellhauer \&  Kroupa 2002).
In particular, theoretical predictions of the correlations between their 
structural and kinematical properties (e.g., central velocity dispersion) 
are generally considered to be important, because such dynamical correlations 
(or ``scaling relations'') for a self-gravitating system are generally 
considered to help discriminate between different formation mechanisms (e.g., 
Djorgovski 1993). In the light of recent discoveries of UCDs and NCs,
it is thus timely and important to discuss  whether such a merger scenario 
(referred to as  ``cluster cannibalism'' in galactic nuclei) is consistent with 
their observed properties.

The purpose of this Letter is to provide the first theoretical predictions on 
the structural and kinematical properties of VMSCs formed by {\it 
dissipationless} multiple cluster merging based on self-consistent numerical 
simulations of nucleus formation. We focus particularly on correlations between 
properties such as luminosity ($L$), effective radius ($R_{\rm e}$), central
velocity dispersion (${\sigma}_{0}$), and surface brightness at $R_{\rm e}$.
The predicted scaling relations combined with current and future observations of 
UCDs (e.g., Drinkwater et al. 2003) and NCs (e.g., B$\ddot{\rm o}$ker et al. 
2004a,b) can provide new insight into the origin of galactic stellar nuclei.
Dissipative formation, which is an alternative formation scenario for VMSCs  
(B$\ddot{\rm o}$ker et al. 2004a,b), will be discussed in  forthcoming papers.

\section{The Model}

We investigate the dynamical evolution of a self-gravitating system 
composed of smaller star clusters (SCs), via numerical simulations carried out 
on a GRAPE board (Sugimoto et al. 1990). Each of the individual SCs that merge 
with one another to form a VMSC, are assumed to have a Plummer density profile 
(e.g., Binney \& Tremaine 1987) with luminosities ($L_{\rm sc}$) and central
velocity dispersions (${\sigma}_{\rm sc}$) consistent with the relation observed 
for GCs (Djorgovski et al. 1997):
\begin{equation}
L_{\rm sc} \propto {{\sigma}_{\rm sc}}^{1.7}. \;
\end{equation}   
The scale length ($a_{\rm sc}$) of a SC is determined by the formula
\begin{equation}
a_{\rm sc} = GM_{\rm sc}/6{{\sigma}_{\rm sc}}^{2}, \;
\end{equation}   
where G and $M_{\rm sc}$ are the gravitational constant
and the mass of the SC, respectively.
Since the-mass-to-light-ratio ($M_{\rm sc}/L_{\rm sc}$) is 
assumed to be constant for all SCs,
$a_{\rm sc}$ and ${\sigma}_{\rm sc}$ are determined 
by the equations (1) and (2) for a given $L_{\rm sc}$ (or $M_{\rm sc}$).
The normalization factor in equation (1) is determined by using 
the observed typical mass ($6\times10^5$ $M_{\odot}$),
the half mass radius (10pc), and the central velocity dispersion  (7 km s$^{-1}$)
for GCs (e.g., Binney \& Tremaine 1987).

A SC system is assumed to have either a uniform disk distribution (referred to 
as ``2D'' models) or a uniform spherical distribution (``3D''), and to be fully 
self-gravitating (i.e., not influenced dynamically by its host galaxy).
For the 2D models, a SC system is assumed to have only rotation initially (no 
velocity dispersion), with its rotational velocity ($V_{\rm rot}$) at a distance 
($r$) from its centre given by: 
\begin{equation}
V_{i}(r)= C_{\rm V}V_{\rm cir}(r), \;
\end{equation}   
where $V_{\rm cir}(r)$ is the circular velocity at $r$ for this system
and $C_{\rm V}$ is a parameter that determines how far the system deviates from 
virial equilibrium (i.e., $0\le C_{\rm V} \le 1$).
For 3D models, a SC system is supported only by its random motions; 
its one-dimensional isotropic dispersion (${\sigma}_{i}$) at radius $r$ is given 
by: 
\begin{equation}
{\sigma}_{i}(r)=C_{\rm V}\sqrt{-\frac{U(r)}{3}},
\end{equation}
where $U(r)$ is the gravitational potential of the system. Here a SC system with 
$C_{\rm V}$ = 1 means that the system is in virial equilibrium (or supported 
fully by rotation). We present the results of models in which all of the SCs are 
within $\sim100$ pc of each other, because they are the most consistent with 
observations of the structural properties of VMSCs.

We investigate two different representative cases: (1)\,the SC system
is composed only of equal-mass SCs (``equal-mass'' case) and, (2)\,the SC system 
is composed of SCs with different masses (``multi-mass'' case). 
For (1), each SC is assumed to have a mass of $2\times 10^6$ $M_{\odot}$
(or $M_{\rm V}$ = $-10.35$ mag) and $a_{\rm sc}$=6.8pc.
For (2), the SCs are assumed to have a luminosity ($M$) function consistent with 
that observed:   
\begin{equation}
\Phi(M)={\rm constant} \times e^{-{(M-M_{0})}^{2}/2{{\sigma}_{m}}^2},
\end{equation}
where $M_{0}(V)$=$-7.27$ mag and ${\sigma}_{m}$=1.25 mag (Harris 1991). 
The number of SCs ($N_{\rm sc}$) in a system is a free parameter, ranging from 2 
to 20 for the equal-mass case and 2 to 200 for the multi-mass one.The model with 
the maximum $N_{\rm sc}$ of 200, corresponds to the most massive VMSCs that have 
been observed (Drinkwater et al. 2003). 
In our simulation, the masses of the stellar particles in the SCs are assumed to 
be equal, so that the total number in the simulation
depends on $N_{\rm sc}$. For example, the particle number 
in the multi-mass model with $N_{\rm sc}$ = 200 is 214858.  
We mainly describe the equal-mass 3D model with $N_{\rm sc}$ = 12
and $C_{\rm V}$ = 0.5 
(the ``fiducial'' model), because this model 
shows both typical behavior of VMSC formation and one of the most
interesting results in the present study.
Also, we describe the (1)\,scaling relations of the simulated VMSCs
and (2)\,parameter dependences of VMSC properties, based on 60 different models.  
The mass, length, time, and velocity units are 2.0 $\times$ $10^6$ $M_{\odot}$,
34.0\,pc, 2.1 $\times$ $10^6$\,yr, and 15.9\,km\,s$^{-1}$.

\placefigure{fig-1}
\placefigure{fig-2}

\section{Results}

As seen from the fiducial model shown in Figure 1, smaller SCs repeatedly merge 
with one another to form bigger clusters through the process of dynamical 
collapse of the SC system. These bigger clusters then merge to form a 
single VMSC with an outer diffuse stellar envelope, within $\sim$ $10^7$\,yr.  
The VMSC has an effective radius ($R_{\rm e}$) of 19.4\,pc
(2.85 $R_{\rm e}$ of the progenitor SC) and within $5R_{\rm e}$ a mass 
of 2.1 $\times$ $10^7$ $M_{\odot}$, which corresponds to $M_{\rm V}$ = $-12.9$ 
mag and is 10.2 times more massive than the mass of the original SCs.
Figure 2 shows the structural and kinematical properties of the VMSC.
Three different non-spherical shapes can be clearly seen in the three projected
mass distributions, which suggests that the VMSC is a triaxial system.
The ellipticity ($\epsilon$), defined as $\epsilon$ = 1-$b/a$, where
$a$ and $b$ are the long and the short axes, respectively, 
in the isodensity contour 
of the projected mass profile, is estimated to be 0.14 for the $x-y$ projection,
0.08 for the $x-z$ projection, and 0.27 for the $y-z$ projection at the 
effective radius of the VMSC. Due to efficient conversion of the initial orbital 
energy of the SCs into internal rotational energy during the SC merging process,
the final VMSC has a non-negligible amount of rotation that
is indicated by moderately high $V_{\rm m}/\sigma_{0}$ ($\sim$ 0.3),
where  $\sigma_{0}$ and $V_{\rm m}$ are the central velocity dispersion
and the maximum line-of-sight velocity of the VMSC  for each projection. 
It should be noted here that a flattened triaxial system with non-negligible 
rotation is remarkably different from typical GCs that have no net rotation and 
quite spherical shapes (mean $\epsilon = 0.07$; e.g., White \& Shawl 1987).

The simulated VMSCs show interesting correlations between their structural and
kinematical parameters (Figure 3). Firstly, more luminous VMSCs have larger 
central velocity dispersions ($\sigma_{0}$) and this correlation can be 
expressed as ${\sigma}_{0}$ $\propto$ $L^{0.31}$, the slope of which is similar 
to that (0.25) of the Faber-Jackson(1976) relation derived for elliptical 
galaxies. Secondly, more luminous VMSCs have larger effective radii 
and the correlation can be expressed as $R_{e}$ $\propto$ $L^{0.38}$,
though the dispersion in this relation is moderately large. This can be compared 
with the corresponding relation ($R_{e}$ $\propto$ $L^{1.06}$) derived for 
elliptical galaxies (Kormendy 1985).  Thirdly, $L$ is more strongly correlated 
with the central surface brightness ($I_{10}$) than the half-light averaged 
surface brightness ($I_{e}$). Although a single line is fitted to all the data 
points for the $L-I_{e}$ relation, it is possible that the relation is different 
over the range $-12$ $\le$ $M_{\rm V}$ $\le$ $-10$\,mag and $-14$ $\le$ $M_{\rm 
V}$ $<$ $-12$\,mag (i.e., there is a hint of V-shaped distribution in Figure 3).

Figure 4 shows the comparison between the locations of all of the simulated 
VMSCs in the $M_{\rm V}-{\sigma}_{0}$ plane and the corresponding observations 
(Drinkwater et al. 2003). Here only (5) UCD points are plotted, since data for 
NCs is not available. The locations of the simulated brighter VMSCs are 
consistent with the observations, and both the simulated and observed data 
points are closer to the Faber-Jackson relation than to the  $M_{\rm V}-
{\sigma}_{0}$ relation of GCs (Djorgovski et al. 1997). This implies that the 
origin of UCDs' structural and kinematically properties is significantly 
different to that of GCs, and is closely associated with physics of multiple 
merging of SCs. Thus the present  results on the scaling relations of VMSCs
clearly show that the scaling relations of VMSCs formed from
multiple SC merging are significantly different both from those 
of their progenitor SCs (or GCs)  and  from those of dynamically
hot early-type galaxies.

The parameter dependences of structural and kinematical properties of the 
simulated VMSCs can be summarized as follows: Firstly VMSCs are likely to be 
more flattened ($\epsilon$ = $0.2 \sim 0.3$) in the 2D models than in the 3D 
models for given parameter values of $C_{\rm V}$ and $N_{\rm sc}$.
Secondly, triaxial VMSCs in some 2D models show large  
$V_{\rm m}/\sigma_{0}$ ($\sim$ 0.4) and {\it figure rotation} like barred 
galaxies. Thirdly, the multi-mass 3D models with large $N_{\rm sc}$ ($\geq$ 50) 
show both smaller $\epsilon$ (i.e., less flattened) and smaller $V_{\rm 
m}/{\sigma}_{0}$ (i.e., less strongly supported by rotation).
This is because a larger number of SCs merge with one another from random 
directions in the multi-mass 3D models. These results suggest that the 
structural and kinematical properties of stellar galactic nuclei (i.e., NCs and 
UCDs) can differ, depending on the merging histories of SCs.

\placefigure{fig-3}
\placefigure{fig-4}

\section{Discussions and conclusions}

We have demonstrated that if VMSCs are formed from the multiple merging
of SCs, with the observed scaling relations of GCs, the scaling relations of 
VMSCs are very different from those of GCs. Ongoing and future photometric and 
spectroscopic observations (e.g., $HST$ ACS and Keck 10m) of the structural and 
kinematical properties of VMSCs will therefore be able to assess the viability 
of  the ``cluster cannibalism'' scenario of stellar nucleus formation. 

We have also shown that (1)\,the intrinsic shapes of VMSCs are more likely to be 
triaxial, and (2)\,some VMSCs can have rotational kinematics. We thus suggest 
that further observations which provide better statistics on: (1)\,the 
$\epsilon$ distributions of the projected isophotal shapes of VMSCs (which 
strongly depend on the intrinsic shapes of VMSCs), and (2)\,the locations of 
VMSCs on the  $\epsilon - V_{\rm m}/{\sigma}_{0}$ plane (which depend on the 
internal kinematics of VMSCs) will help discriminate between different VMSC 
formation scenarios. 

Our study also has other important implications. Firstly, the significant amount 
of rotation observed for the metal-poor populations in the most massive Galactic 
globular cluster $\omega$ Cen (e.g., Norris et al. 1997), provides evidence 
that $\omega$~Cen may have originated from the nucleus of a dwarf galaxy. A 
growing number of observations and theoretical studies have recently suggested 
that $\omega$~Cen was previously a nucleus of a ancient nucleated dwarf orbiting 
the young Galaxy (e.g., Hilker \& Richtler 2000, 2002; Bekki \& Freeman 2003).  
The present study has demonstrated that VMSCs formed from SC merging can have a 
significant amount of rotation (i.e., larger $V_{\rm m}/{\sigma}_{0}$), due to 
the conversion of orbital energy into intrinsic rotational energy. Therefore the 
rotational kinematics of the metal-poor populations of $\omega$~Cen reflect the 
past merging of SCs {\it in the central region of its host galaxy}.

Secondly, we can significantly underestimate the mass-to-light-ratios ($M/L$)
of VMSCs, in particular those with large $V_{\rm m}/{\sigma}_{0}$, if we 
estimate the $M/L$ by adopting the commonly used formula in which 
$M/L$ is derived only from the central velocity dispersion and effective radius 
(e.g., Meylan et al. 2001). This is simply because the implicit assumption (in 
the formula) that kinematical energy can be accurately measured by the central 
velocity dispersion ${\sigma}_{0}$ alone is not valid for these VMSCs. 
Accordingly, the real values of $M/L$ for UCDs may be even larger than the 
moderately large values ($M/L =2$--4) that are observed (Drinkwater et al. 2003), 
if UCDs indeed have rotational kinematics -- something which future 
spectroscopic observations of UCDs can confirm.

Thirdly, the observed young stellar populations in NCs (e.g., B$\ddot{\rm o}$ker 
et al. 2004b) can be due to nuclear star formation triggered by dynamical 
interaction between a NC and its surrounding interstellar gas. Recent numerical 
simulations have suggested that dynamical interaction between self-gravitating 
triaxial systems with figure rotation and the surrounding gas can cause rapid 
gas transfer to the central region and trigger subsequent starbursts there 
(Bekki \& Freeman 2002). The present study has shown that some triaxial VMSCs
have figure rotation (particularly when they are formed from SCs with initial 
disky distributions). Therefore, we suggest that the observed young populations 
in NCs may be due to past dynamical interaction between the triaxial NCs with 
figure rotation and the surrounding gas.

Although we have not discriminated between UCDs and NCs throughout this paper, 
there exists a possible remarkable difference in the size distribution between 
the two: The observed effective radii range from 15\,pc to 30\,pc for UCDs 
and from 2.4\,pc to 5\,pc for NCs. This difference can possibly be ascribed to 
the young stellar populations with compact spatial distributions within NCs
(i.e., the size distributions for old populations of NCs could not be so 
different from those of UCDs). However, it is equally possible that the 
difference is due to the differences in formation processes between UCDs and NCs. 
Differences in scaling relations between UCDs and NCs, if any, would tell us 
something about the origin of the difference between UCDs and NCs. 

\acknowledgments
K.B. and W.J.C. acknowledge the financial support of the Australian Research 
Council throughout the course of this work.

\clearpage


\newpage
\plotone{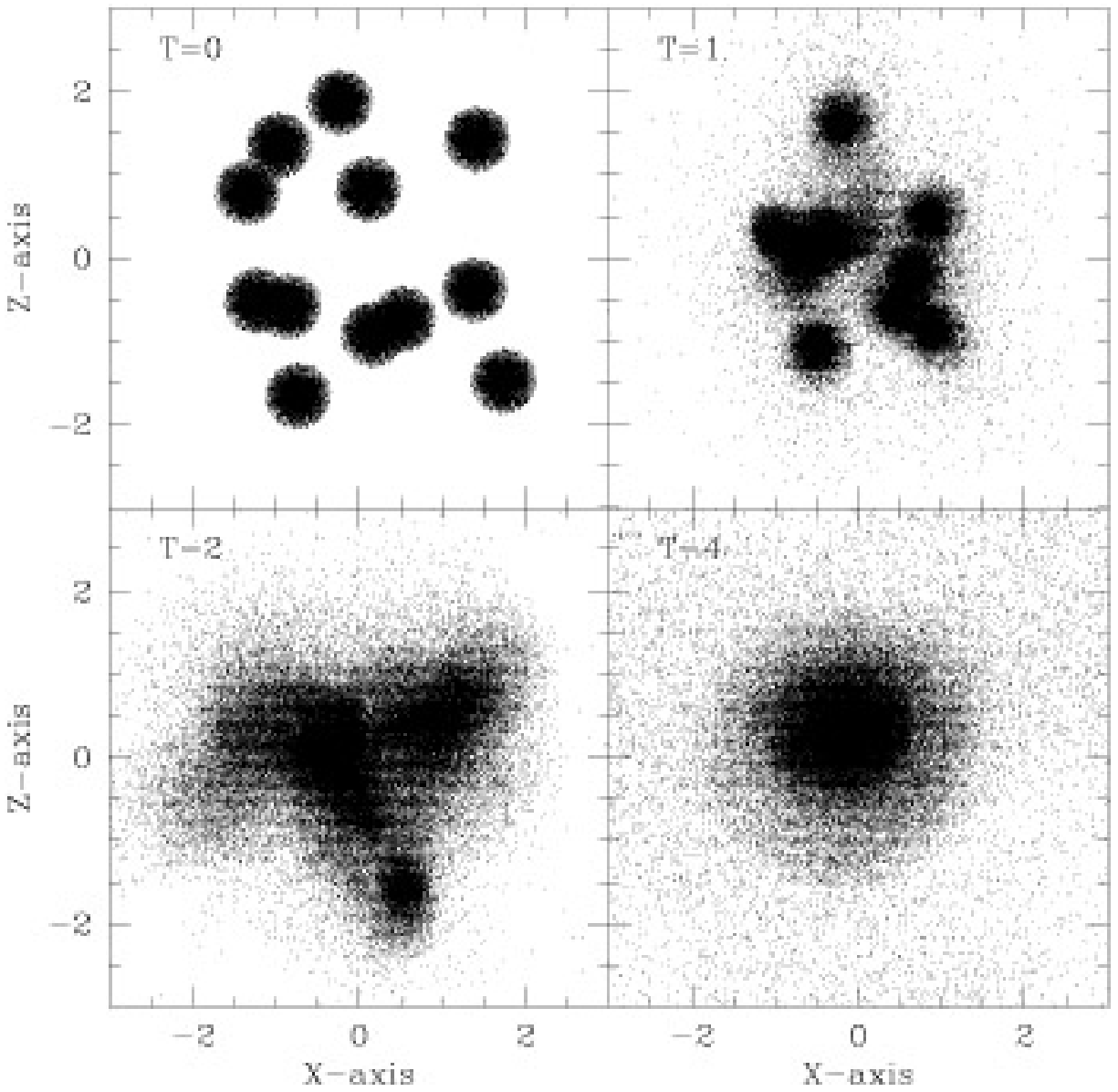}
\figcaption{
Morphological evolution of twelve equal-mass star clusters (SCs)
projected onto the $x-z$ plane for the fiducial model
(with an initial total mass of 2.4 $\times$ $10^7$ $M_{\odot}$).
The length and the time are given in our units (34.0\,pc and 
2.1 $\times$ $10^6$\,yr, respectively). Here the time $T$ represents 
the time that has elapsed since the start of the simulation.
\label{fig-1}}

\newpage
\plotone{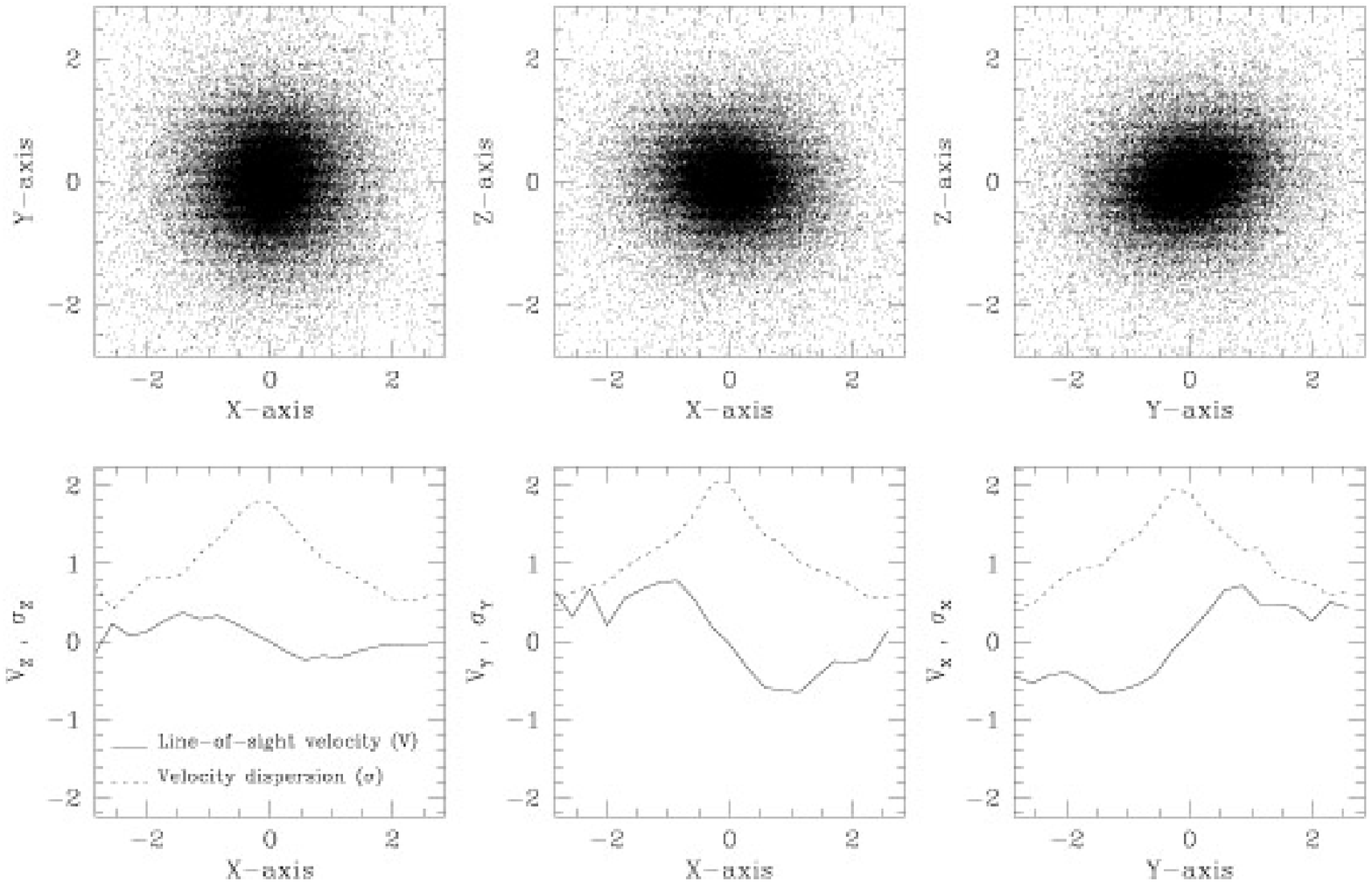}
\figcaption{
Final mass distributions ({\it upper}) and kinematical properties ({\it lower}), 
projected onto the $x-y$ plane ({\it left}), the $x-z$ plane ({\it middle}),
and $y-z$ plane ({\it right}) in the fiducial model at $T = 32.0$ in our units 
(corresponding to 6.7 $\times$ $10^7$\,yr). {\it Solid} and {\it dotted} lines 
in each lower panel represent the radial profile of line-of-sight velocity
and that of the velocity dispersion, respectively. The length and the velocity 
are given in our units (34.0\,pc and 15.9\,km\,s$^{-1}$, respectively). 
\label{fig-2}}

\newpage
\plotone{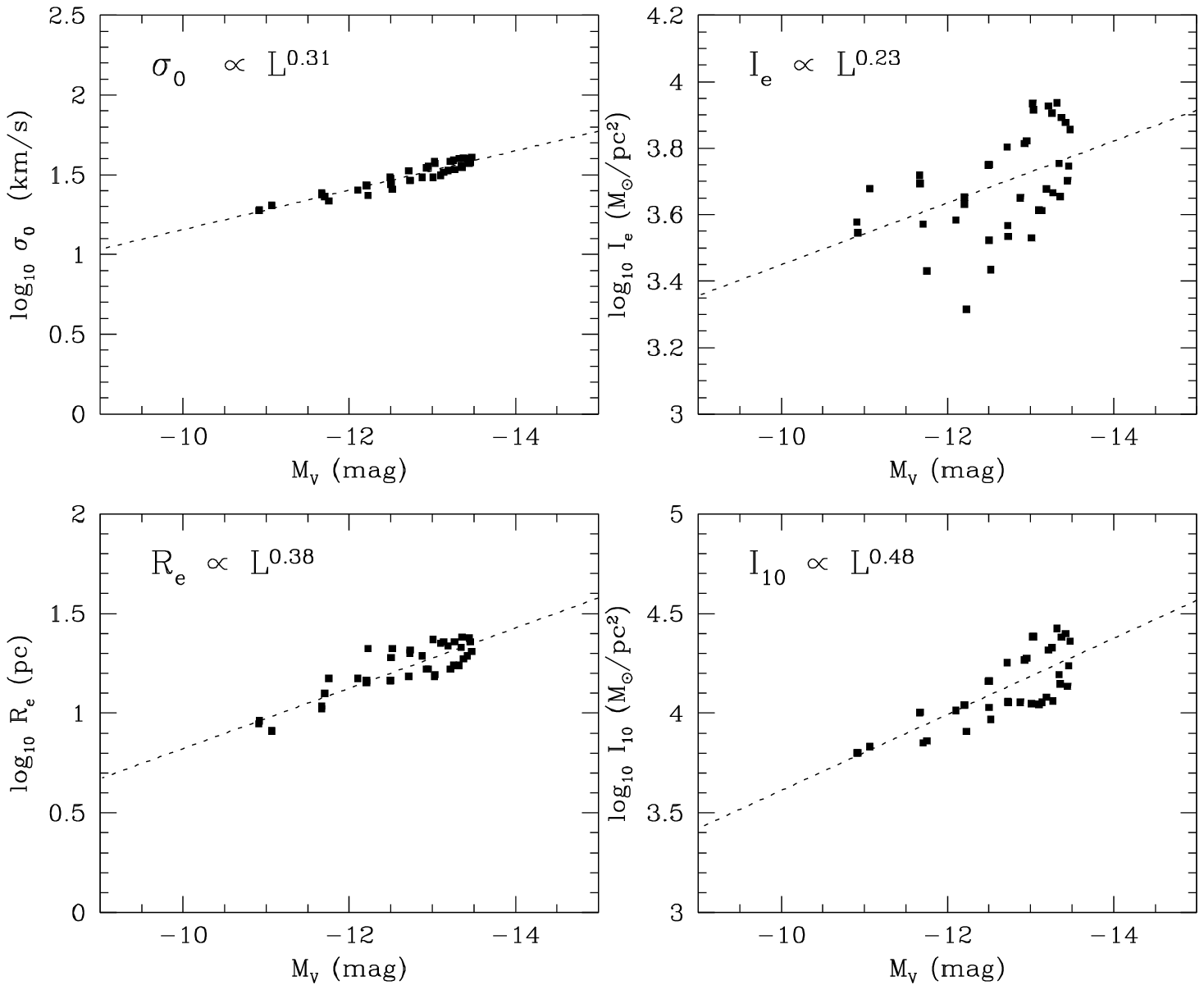}
\figcaption{
Correlations of structural and kinematical parameters with $M_{\rm V}$ ($V-$band 
absolute magnitude) for the VMSCs in 40 models including equal-mass 2D and 3D 
models with $C_{\rm V}$ =0 and 0.5. Projected central velocity dispersion 
($\sigma_{0}$; {\it upper left}), half-light-averaged surface brightness 
($I_{e}$; {\it upper right}), effective radius ($R_{e}$; {\it lower left}),
and central surface brightness ($I_{10}$; {\it lower right}) are plotted against 
$M_{\rm V}$. Here the central surface brightness $I_{10}$ is expressed as  
$0.1L/\pi/{R_{10}}^{2}$, where $L$, $R_{10}$ are the total luminosity of a VMSC
and the radius within which 10\% of $L$ is included, respectively. The best fit 
scaling relation for the VMSCs is derived for  each panel using the least square 
fitting method and described as a {\it dotted} line with the derived relation 
(e.g., ${\sigma}_{0}$ $\propto$ $L^{0.31}$). 
\label{fig-3}}

\newpage
\plotone{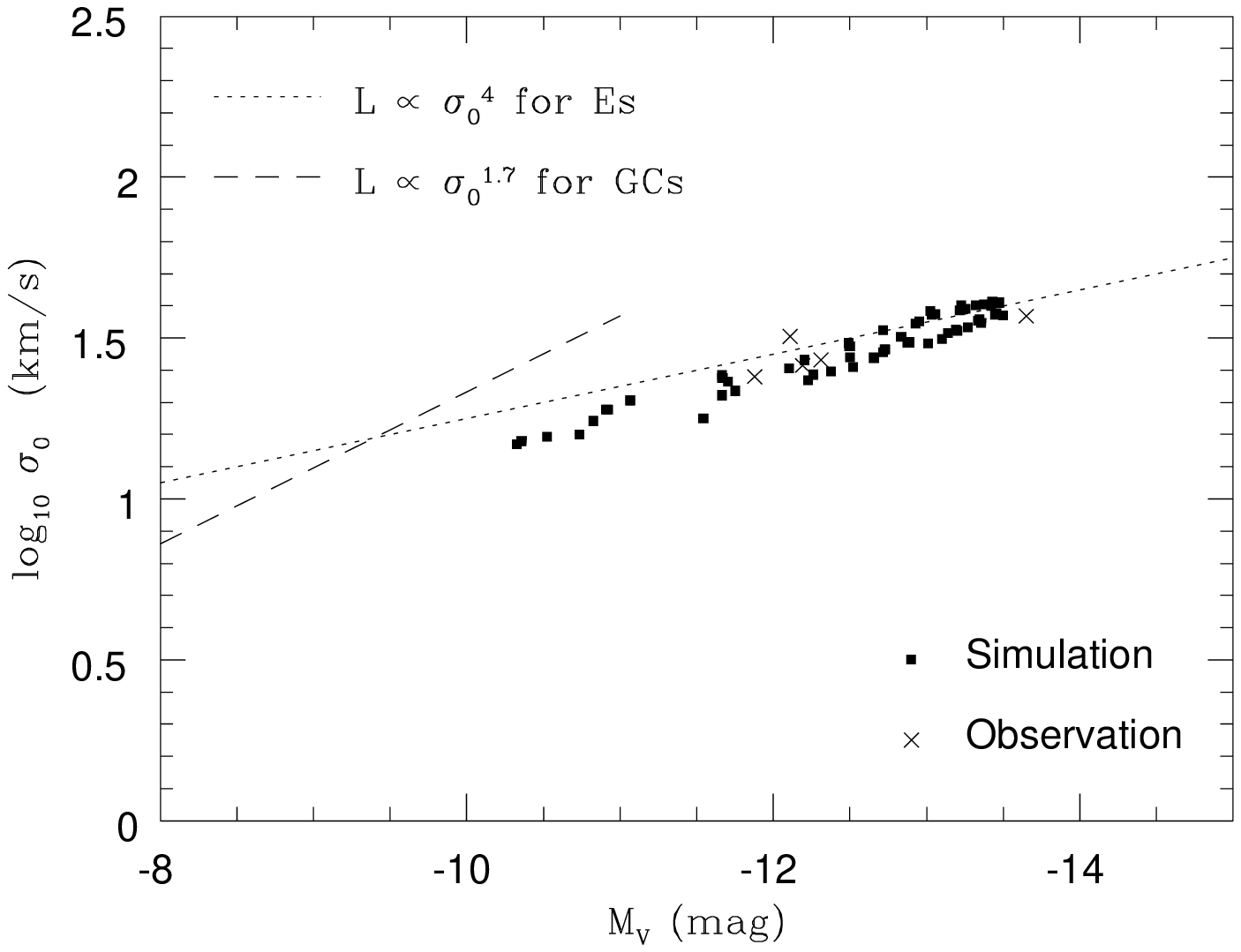}
\figcaption{
Correlations of ${\sigma}_{0}$  with $M_{\rm V}$ for the simulated VMSCs ({\it 
filled squares}) and the observations ({\it crosses}). The results of 60 models,  
including both equal-mass 2D/3D models and multi-mass ones, are shown. Only 5 
UCDs with known ${\sigma}_{0}$  (Drinkwater et al. 2003) are plotted (no 
velocity dispersion data for NCs are available). For comparison, the observed 
relations are given by a {\it dashed} line for GCs (Djorgovski et al. 1997)
and a {\it dotted} line for Es (Faber \& Jackson 1976).
\label{fig-4}}


\end{document}